\documentclass[iop]{emulateapj}

\usepackage{epstopdf}
\usepackage{epsfig}
\usepackage{graphics}
\usepackage{graphicx}
\usepackage{natbib}
\begin{document}

\title{Modeling Multi-Wavelength Stellar Astrometry. II. Determining Absolute Inclinations, Gravity Darkening Coefficients, and Spot Parameters of Single Stars with SIM Lite}
\author{Jeffrey L. Coughlin\altaffilmark{1,3}, Thomas E. Harrison\altaffilmark{1}, \& Dawn M. Gelino\altaffilmark{2}}
\altaffiltext{1}{Department of Astronomy, New Mexico State University, P.O. Box 30001, MSC 4500, Las Cruces, New Mexico 88003-8001; jlcough@nmsu.edu}
\altaffiltext{2}{NASA Exoplanet Science Institute, California Institute of Technology, Pasadena, CA 91125}
\altaffiltext{3}{NSF Graduate Research Fellow}

\begin{abstract}

We present a novel technique to determine the absolute inclination of single stars using multi-wavelength sub-milliarcsecond astrometry. The technique exploits the effect of gravity darkening, which causes a wavelength-dependent astrometric displacement parallel to a star's projected rotation axis. We find this effect is clearly detectable using SIM Lite for various giant stars and rapid rotators, and present detailed models for multiple systems using the \textsc{reflux} code. We also explore the multi-wavelength astrometric reflex motion induced by spots on single stars. We find that it should be possible to determine spot size, relative temperature, and some positional information for both giant and nearby main-sequence stars utilizing multi-wavelength SIM Lite data. This data will be extremely useful in stellar and exoplanet astrophysics, as well as supporting the primary SIM Lite mission through proper multi-wavelength calibration of the giant star astrometric reference frame, and reduction of noise introduced by starspots when searching for extrasolar planets.

\end{abstract}


\section{Introduction}
\label{introsection}

SIM Lite is currently expected to have $\sim$80 spectral channels \citep{SIM09}, spanning 450 to 900 nm, thus allowing multi-wavelength microarcsecond astrometry, which no current or planned ground or space-based astrometric project, (GAIA, CHARA, VLT/PRIMA, etc.) is able to match. We showed in our first paper \citep{Coughlin10}, hereafter referred to as Paper I, the implications multi-wavelength microarcsecond astrometry has for interacting binary systems. In this paper, we discuss an interesting effect we encountered while modeling binary systems, namely that gravity darkening in stars produces a wavelength dependent astrometric offset from the center of mass that increases with decreasing wavelength. It is possible to use this effect to derive both the inclination and gravity darkening exponent of a star in certain cases.

Determining the absolute inclination of a given star has many practical applications. There is much interest in the formation of binary stars, where whether or not the spin axis of each star is aligned with the orbital axis provides insight into the formation history of the system \citep{Turner95}. The mutual inclination between the stellar spin axes and orbital axis can greatly affect the rate of precession, which is used to probe stellar structure and test general relativity \citep{Sterne39a,Sterne39b,Sterne39c,Kopal59,Jeffery84}. \citet{Albrecht09} recently reconciled a 30-year-old discrepancy between the observed and predicted precession rate of DI Herculis through observations which showed the stellar spin axes were nearly perpendicular to the orbital axis. Along similar lines, extrasolar planets discovered via the radial velocity technique only yield the planetary mass as a function of the inclination of the orbit \citep{Mayor95, Noyes97, Marcy00}, and thus, if one assumes the planetary orbit and stellar rotation axes are nearly parallel, determining the absolute inclination of the host star yields the absolute mass of the planet. If the stellar spin axis is found not to be parallel to the planetary orbital axis, this provides valuable insights into the planet's formation, migration, and tidal evolution histories \citep{Winn06,Fabrycky09}. A final example is the study of whether or not the spin axes of stars in clusters are aligned, which both reveals insight into their formation processes, as well as significantly affects the determination of the distances to those clusters \citep{Jackson10}.

Our proposed technique can also be used in conjunction with other methods of determining stellar inclination to yield more precise inclination values and other stellar parameters of interest. \citet{Gizon03} and \citet{Ballot06} have shown that one can derive the inclination of the rotation axis for a given star using the techniques of astroseismology given high-precision photometry with continuous coverage over a long baseline, such as that provided by the CoRoT and Kepler missions. This technique is sensitive to rotation rates as slow as the Sun's, but becomes easier with faster rotation rates. \citet{Domiciano04} discuss how spectro-interferometry can yield both the inclination angle and amount of differential rotation for a star, parameterized by $\alpha$. For both eclipsing binaries and transiting planets, the observation of the Rossiter-McLaughlin (RM) effect can yield the relative co-inclination between the two components \citep{Winn06,Albrecht09,Fabrycky09}. The technique we propose in this paper would be complementary to these techniques in several ways. First, it would provide an independent check on the derived inclination axis from each method, confirming or refuting the astroseismic models and spectro-interferometric and RM techniques. Second, in principle the astroseismic technique is not dependent on the gravity darkening coefficient $\beta_{1}$, and the spectro-interferometric technique is correlated with the value for $\alpha$; combining techniques would yield direct and robust observationally determined values for $i$, $\alpha$, and $\beta_{1}$. Finally, the accurate, observational determination of $\alpha$ and $\beta_{1}$, (along with stellar limb-darkening), is critical to accurately deriving the co-inclination from the RM effect, as well as other quantities in stellar and exoplanet astrophysics.

In this paper, we also present models for and discuss the determination of spot location, temperature, and size on single stars, which produce a wavelength-dependent astrometric signature as they rotate in and out of view. Star spots are regions on the stellar surface where magnetic flux emerges from bipolar magnetic regions, which blocks convection and thus heat transport, effectively cooling the enclosed gas, and thus are fundamental indicators of stellar magnetic activity and the internal dynamos that drive it. \citet{Isik07} discuss how the observation of spot location, duration, stability, and temperature can probe the stellar interior and constrain models of magnetic flux transport. Through the observation of the rotation rates of starspots at varying latitudes, one is able to derive the differential rotation rate of the star \citep{Collier02}, which may be directly related to the frequency of starspot cycles. Mapping spots in binary star systems provides insight into the interaction between the magnetic fields of the two components, which can cause orbital period changes \citep{Applegate92}, radii inflation \citep{Lopez-Morales07,Morales08}, and may possibly explain the $\sim$2-3 hour period gap in cataclysmic variable systems \citep{Watson07}. Detecting and characterizing star spots via multi-wavelength astrometry would be complementary to other existing techniques, namely optical interferometry \citep{Wittkowski02}, tomographic imaging \citep{Donati06,Auriere08}, photometric monitoring \citep{Alekseev04,Mosser09}, and in the future, microlensing \citep{Hwang10}.

We present the details of our modeling code, \textsc{reflux}, in \S\ref{refluxsection}, discuss the inclination effect and present models for multiple stars in \S\ref{incsection}, discuss the spot effects and present models in \S\ref{spotsection}, and present our conclusions in \S\ref{conclusionsection}.

\section{The {\sc reflux} Code}
\label{refluxsection}

\textsc{reflux}\footnotemark[1] is a code that computes the flux-weighted astrometric reflex motion of binary systems. We discussed the code in detail in Paper I, but in short, it utilizes the Eclipsing Light Curve (ELC) code, which was written to compute light curves of eclipsing binary systems (Orosz \& Hauschildt 2000). The ELC code represents the surfaces of two stars as a grid of individual luminosity points, and calculates the resulting light curve given the provided systemic parameters. ELC includes the dominant physical effects that shape a binary's light curve, such as non-spherical geometry due to rotation, gravity darkening, limb darkening, mutual heating, reflection effects, and the inclusion of hot or cool spots on the stellar surface. For the work in this paper we have simply turned off one of the stars, thus allowing us to probe the astrometric effects of a single star. To compute intensity, ELC can either use a blackbody formula or interpolate from a large grid of NextGen model atmospheres (Hauschildt et al. 1999). For all the simulations in this paper, we have used the model atmosphere option, and will note now, and discuss more in detail later, that the calculation of limb-darkening is automatically included in NextGen model atmospheres. These artificially derived limb-darkening coefficients have recently been shown to be in error by as much as $\sim$10-20\% in comparison to observationally derived values \citep{Claret08}, and thus their uncertainties must be included, although for this work, due to symmetry, we find the introduced error is negligible. For all our simulations, we model the U, B, V, R, I, J, H, and K bands for completeness and comparison to future studies, though we note that SIM Lite will not be able to observe in the U, J, H, or K bandpasses.

\footnotetext[1]{\textsc{reflux} can be run via a web interface from \url{http://astronomy.nmsu.edu/jlcough/reflux.html}. Additional details as to how to set-up a model are presented there.}

\section{Inclination and Rotation}
\label{incsection}

\begin{figure*}
\centering
\begin{tabular}{cc}
\epsfig{file=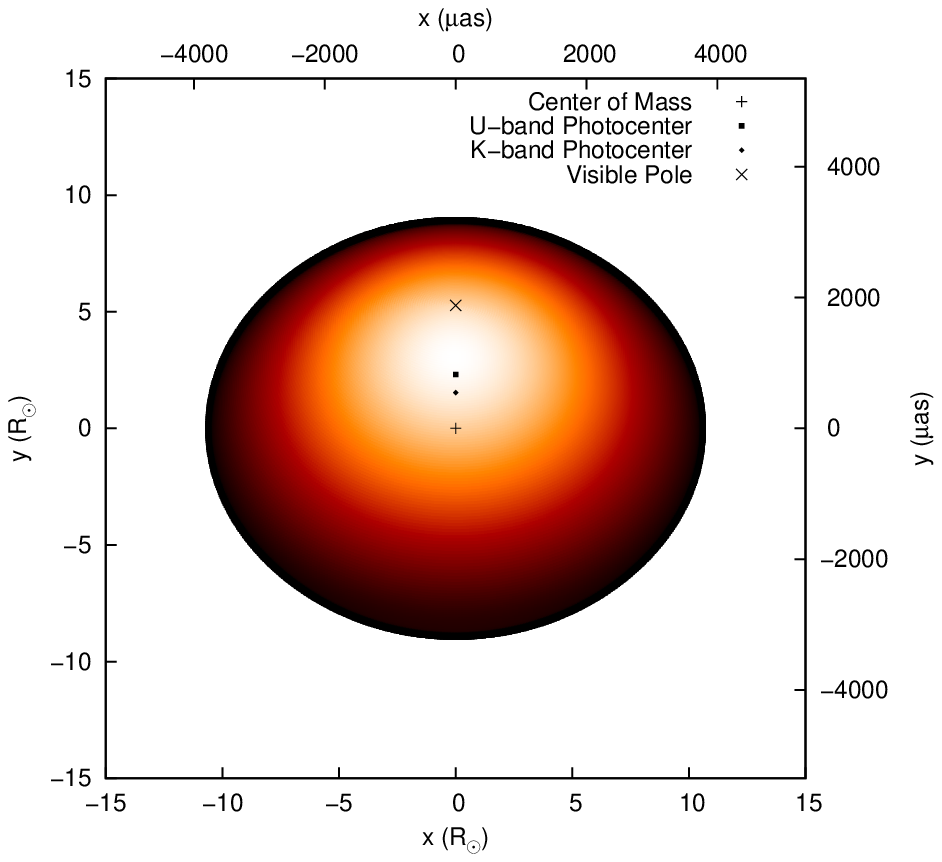, width=0.474\linewidth} &
\epsfig{file=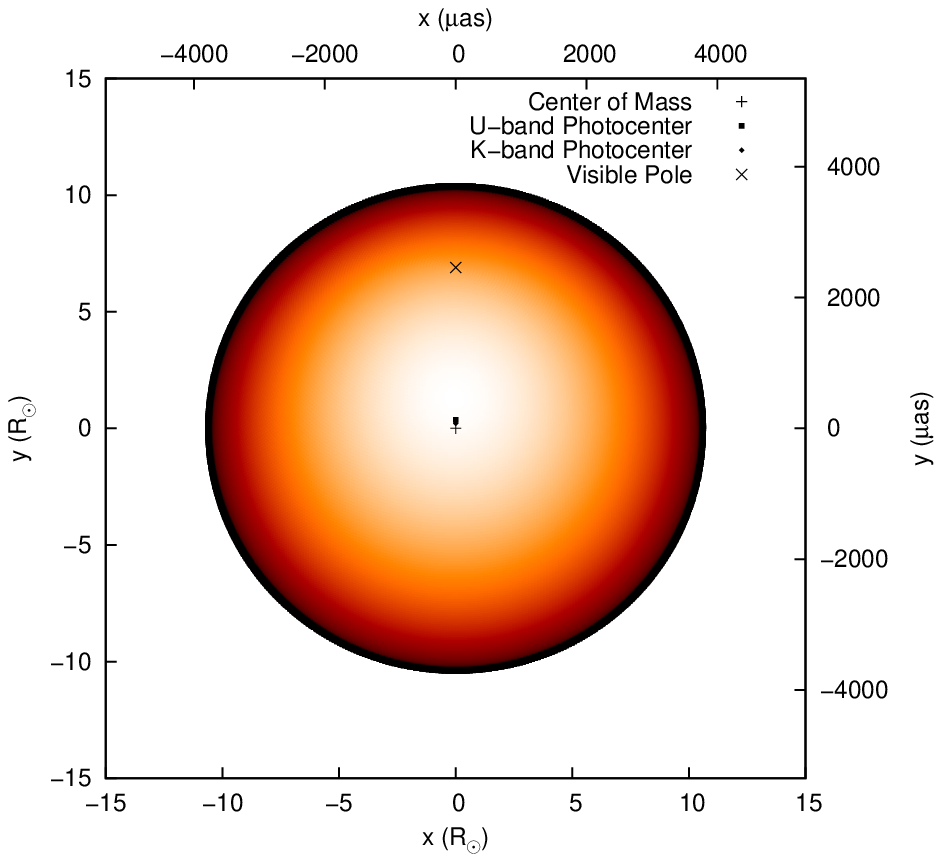, width=0.474\linewidth} \\
\end{tabular}
\caption{Illustration of the inclination effect using brightness maps of Capella Ab, which has an inclination of 42.788$\degr$ \citep{Torres09}. Left: Capella Ab artificially spun-up to near its break-up speed, to accentuate the gravity darkening effect for ease of viewing. As can be seen, the photocenter of the system is dramatically shifted away from the center of mass, towards the visible pole, which is brighter than the rest of the star due to gravity darkening. Furthermore, since the pole is physically hotter, the U-band photocenter is shifted more than the K-band photocenter, and in the direction of the projected rotation axis, or y-axis. Right: Capella Ab at its actual rotation period. As can be seen, the actual effect is small compared to the angular size of the star on the sky, but still large compared to the 1 $\mu$as benchmark of SIM Lite. The presence of gravity darkening is clearly visible, which causes a decrement in flux towards the limb of the star. Note that the broad wavelength covereage of SIM Lite will only cover the B, V, R, and I bandpasses.}
\label{incfig}
\end{figure*}

The astrophysical phenomenon of gravity darkening, also sometimes referred to as gravity brightening, is the driving force behind the ability to determine the inclination of a single star using multi-wavelength astrometry. A rotating star is geometrically distorted into an oblate spheroid, such that its equatorial radius is greater than its polar radius, and thus the poles have a higher surface gravity, and the equator a lower surface gravity, than a non-rotating star with the same mass and average radius. This increased surface gravity, $g$, at the poles results in a higher effective temperature, T$_{\rm eff}$, and thus luminosity; decreased $g$ at the equator results in a lower T$_{\rm eff}$ and luminosity. This temperature and luminosity differential causes the star's center of light, or photocenter, to be shifted towards the visible pole, away from the star's gravitational center of mass. Since the inclination determines how much of the pole is visible, the amount of displacement between the photocenter and the center of mass is directly related to the inclination. Furthermore, since the luminosity difference effectively results from a ratio of blackbody luminosities of differing temperatures, the effect is wavelength dependent, with shorter wavelengths shifted more than longer wavelengths. Thus, the amount of displacement between the measured photocenter in two or more wavelengths is directly related to the inclination. See Figure~\ref{incfig} for an illustration of the effect.

An additional complicating factor is the exact dependence of temperature on local gravity. \citet{Zeipel24} was the first to derive the quantitative relationship between them, showing that $T_{\rm eff}^{4} \propto g^{\beta_{1}}$, where $\beta_{1}$ is referred to as the gravity darkening exponent. The value of $\beta_{1}$ has been a subject of much study and debate; for a complete review, see \citet{Claret00}, who presents both an excellent discussion of past studies, as well as new, detailed computations of $\beta_{1}$ using modern models of stellar atmospheres and internal structure that encompass stars from 0.08 to 40 M$_{\sun}$. Since the value of $\beta_{1}$ affects the temperature differential between equator and pole, the multi-wavelength displacement will also be dependent on the value of $\beta_{1}$. The total amplitude of the effect will be scaled by the angular size of the star, which depends on both its effective radius and distance. Thus, in total, the components of this inclination effect are the effective stellar radius, distance, effective temperature, rotation rate, $\beta_{1}$, and inclination of the star. In principle, one is able to determine the effective stellar radius, effective temperature, rotation rate, and distance of a target star using ground-based spectroscopy and space-based parallax measurements, including from SIM Lite. Thus, when modeling the multi-wavelength displacement of the stellar photocenter, the only two components that need to be solved for are the inclination and $\beta_{1}$, with $\beta_{1}$ already having some constraints from theory.

A good trio of stars for modeling and testing this inclination effect are the components of the binary system Capella, (Aa and Ab), and the single star Vega. \citet{Torres09} has very recently published an extremely detailed analysis of both the binary orbit of Capella and the physical and evolutionary states of the individual components, providing both new observations, as well as drawing from the previous observations and analyses of \citet{Hummel94} and \citet{Strassmeier01}. Vega, in addition to being one of the most well-studied stars in the sky, has recently been discovered to be a very rapid rotator seen nearly pole-on \citep{Aufdenberg06,Peterson06,Hill10}. In total, these three stars represent both slow and rapid rotators for giant and main-sequence stars at a range of temperatures, as Capella Aa is a slow-rotating K-type giant, Capella Ab is a fast-rotating G-type giant, and Vega is a very fast-rotating A-type main-sequence star. With many ground-based interferometric observations to compare with, and being bright and nearby, these stars also present excellent targets for SIM Lite.

\begin{figure}
\centering
\epsfig{file=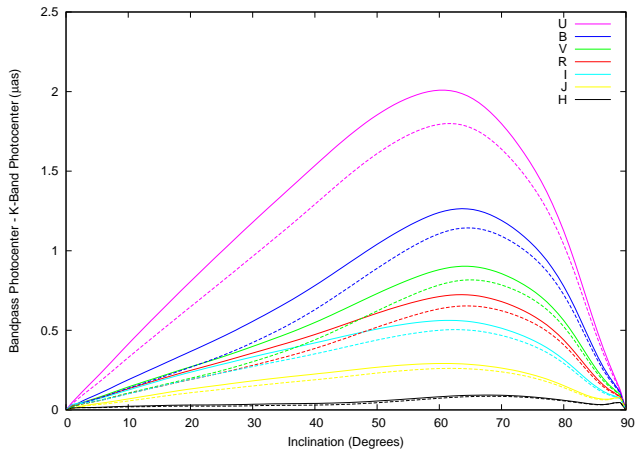, width=\linewidth}
\caption{The astrometric displacement of each bandpass with respect to \emph{K}-band versus inclination for Capella Aa. Dashed lines are a model with $\beta _{1}$ decreased by 10\%. Due to the slow rotation rate of Capella Aa, the effect is limited to a maximum of $\sim$2.0 microarcseconds between U and K-band, and only a maximum of $\sim$0.7 $\mu$as between B and I-band, where SIM Lite will operate. This puts the detection of this effect for Capella Aa at the very edge of SIM Lite's capability.}
\label{CapellaAaFig}
\end{figure}

\begin{figure}
\centering
\epsfig{file=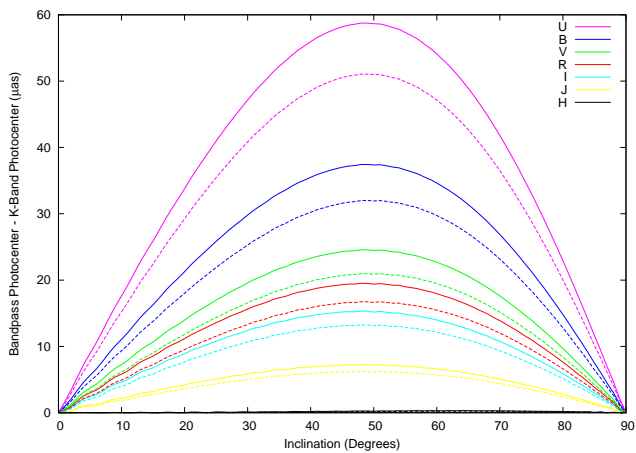, width=\linewidth}
\caption{The astrometric displacement of each bandpass with respect to \emph{K}-band versus inclination for Capella Ab. Dashed lines are a model with $\beta_{1}$ decreased by 10\%. Due to the fast rotation rate of a Capella Ab like star, the effect is moderate with tens of microarcseconds of displacement, and thus these types of stars are excellent targets for SIM Lite. The actual inclination of Capella Ab is 42.788$\degr$ \citep{Torres09}, and thus Capella Ab itself should show a large shift of the photocenter with wavelength. Note that the broad wavelength covereage of SIM Lite will only cover the B, V, R, and I bandpasses.}
\label{CapellaAbFig}
\end{figure}

\begin{figure}
\centering
\epsfig{file=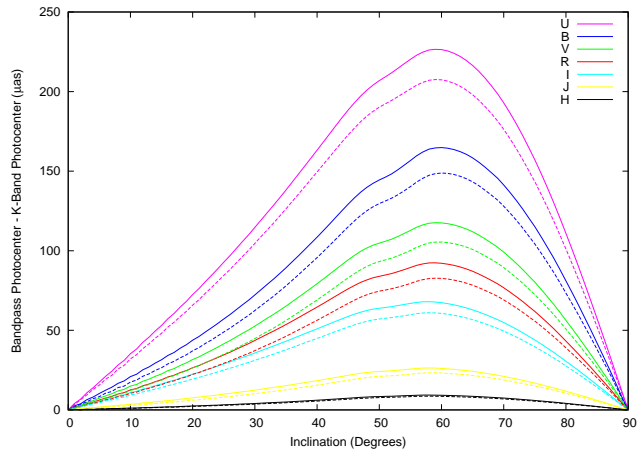, width=\linewidth}
\caption{The astrometric displacement of each bandpass with respect to \emph{K}-band versus inclination for Vega. Dashed lines are a model with $\beta_{1}$ decreased by 10\%. Note that due to the very fast rotation of Vega, along with a high value of $\beta_{1}$, the effect can be quite large, at a couple hundred micro-arcseconds. For a Vega-like star, SIM Lite observations would yield very accurate values for $\beta_{1}$ and the inclination. For Vega itself, which is known to be nearly pole on, with an inclination of 5.7$\degr$ \citep{Hill10}, there should be a B-band minus I-band displacement of 6.0 $\mu$as, still detectable by SIM Lite. Note that the broad wavelength covereage of SIM Lite will only cover the B, V, R, and I bandpasses.}
\label{VegaFig}
\end{figure}

We use the \textsc{reflux} code to generate models of the astrometric displacement from \emph{U}-band to \emph{H}-band, with respect to the \emph{K}-band photocenter, for inclinations from 0 to 90$\degr$, for each star, as shown in Figures~\ref{CapellaAaFig},~\ref{CapellaAbFig}, and~\ref{VegaFig}. We use systemic parameters given by \citet{Torres09} for Capella Aa and Ab, and by \citet{Aufdenberg06} and \citet{Peterson06} for Vega, listed in Tables~\ref{CapellaAaTable},~\ref{CapellaAbTable}, and~\ref{VegaTable} respectively. We employ the model atmospheres incorporated into the ELC code, as well as automatically chosen values for $\beta_{1}$ based on Figure 1 of \citet{Claret00}. Additionally, in each figure we show a dashed line to indicate the effect of decreasing the gravity darkening coefficient by 10\% to simulate the uncertainty of the models \citep{Claret00} and explore the correlation with other parameters.

\begin{deluxetable}{lc}
\tablewidth{0pt}
\tablecaption{Parameters for Capella Aa}
\tablecolumns{2}
\tablehead{Parameter & Value\tablenotemark{a}}
\startdata
Distance (pc) & 12.9\\
Rotation Period (Days) & 106.0\\
Mass (M$_{\sun}$) & 2.70\\
Radius (R$_{\sun}$) & 12.2\\
Effective Temperature (K) & 4940\\
$\beta_{1}$ & 0.43
\enddata
\label{CapellaAaTable}
\tablenotetext{a}{Values from \citet{Torres09}}
\end{deluxetable}

\begin{deluxetable}{lc}
\tablewidth{0pt}
\tablecaption{Parameters for Capella Ab}
\tablecolumns{2}
\tablehead{Parameter & Value\tablenotemark{a}}
\startdata
Distance (pc) & 12.9\\
Rotation Period (Days) & 8.64\\
Mass (M$_{\sun}$) & 2.56\\
Radius (R$_{\sun}$) & 9.2\\
Effective Temperature (K) & 5700\\
$\beta_{1}$ & 0.39
\enddata
\label{CapellaAbTable}
\tablenotetext{a}{Values from \citet{Torres09}}
\end{deluxetable}

\begin{deluxetable}{lc}
\tablewidth{0pt}
\tablecaption{Parameters for Vega}
\tablecolumns{2}
\tablehead{Parameter & Value\tablenotemark{a}}
\startdata
Distance (pc) & 7.76\\
Rotation Period (Days) & 0.521\\
Mass (M$_{\sun}$) & 2.11\\
Radius (R$_{\sun}$) & 2.5\\
Effective Temperature (K) & 9602\\
$\beta_{1}$ & 1.02
\enddata
\label{VegaTable}
\tablenotetext{a}{Values from \citet{Aufdenberg06} and \citet{Peterson06}}
\end{deluxetable}

As can be seen from these models, we find that the effect is quite large for a Capella Ab-like or Vega-like fast rotator, but only marginally detectable for a slower-rotating system like Capella Aa. This also implies that this effect would not be detectable for a slow-rotating, main-sequence star like our Sun. Our modeling confirms this, showing a total U-K amplitude of $\ll$ 0.1 $\mu$as for a 1.0 M$_{\sun}$, 1.0 R$_{\sun}$ star with a rotation period of 30.0 days at 10.0 parsecs. These conclusions on detectability are made with the assumption that, for bright stars like these, SIM Lite can achieve its microarcsecond benchmark. We show this is possible in narrow angle (NA) mode by employing the SIM Differential Astrometry Performance Estimator (DAPE) \citep{Plummer09}. For a target star with magnitude V$=$5, and a single comparison star with V$=$10 located within a degree of it on the sky, by integrating 15 seconds on the target, and 30 seconds on the reference, for 10 visits at 5 chop cycles each, a final precision of $\pm$1.01 $\mu$as is achieved in only 1.04 hours of total mission time. For a fainter target with $V$=10, this precision is only reduced to $\pm$1.32 $\mu$as in the same amount of mission time. In utilizing NA mode, one must be careful in choosing the reference star(s), to ensure that they are not stars with a substantial wavelength dependant centroid. Given the only constraints on reference stars are that they need to have V $\gtrsim$ 10 and are within one degree on the sky, one could easily choose a slow-rotating, main-sequence star, determined as such via ground-based observations, as a wavelength-independent astrometric reference star. We also note that wide angle SIM Lite measurements, with a precision of $\sim$5 $\mu$as, may not detect the wavelength dependent photoceter of a system like Capella, but will have no difficulty detecting it in stars like Capella Ab or Vega.

The effect of decreasing the gravity darkening exponent is to decrease the total amplitude of the effect in each wavelength, with shorter wavelengths affected more than longer wavelengths. Thus, the choice of gravity darkening exponent is intimately tied to the derived inclination. If one were to model observed data with a gravity darkening exponent that was $\sim$10\% different than the true value, they would derive an inclination that would also be $\sim$10\% different from the true inclination. However, the two combinations of inclination and gravity darkening exponent do not produce identical results, and can be distinguished with a sufficient precision at a number of wavelengths. For example, if one were to adopt the nominal value for $\beta_{1}$ and derive an inclination of 40 degrees for a Vega-like star, then adopt a $\beta_{1}$ value that was 10\% lower, one would derive an inclination of 43 degrees, a 7.5\% change. In this case though, with the lower $\beta_{1}$ value, the measured photocenter in the U, B, V, R, I, J, and H bandpasses, with respect to the K-band photocenter, would differ from the nominal $\beta_{1}$ model by $\sim$0.5, -1.0, -2.0, -2.0, -1.6, -1.0, and 0.2 $\mu$as respectively. Note that for B, V, R, and I, where SIM Lite can observe, these discrepancies, on the order of $\sim$1.0 $\mu$as, should be large enough to be distinguished in NA mode. Thus, a unique solution exists for the values of \emph{i} and $\beta_{1}$ if the photocenter is measured in three or more wavelengths. (The photocenter of one wavelength is used as a base measurement that the photocenters of other wavelengths are measured with respect to, as we have chosen K-band as the base measurement in our models. With the photocenter measured in three or more wavelengths total, there are two or more photocenter difference measurements, with two unknown variables for which to solve.) Another complication is the possibility of having equally good fitting high and low solutions for $\emph{i}$. For example, if one observed and determined a best-fit inclination of 70 degrees for a Vega-like star, one could obtain a reasonably good fit as well at 46 degrees, (see Fig~\ref{VegaFig}). However, just as in the case of the uncertainity in the value of $\beta_{1}$, discernible discrepancies would exist. In this case, the discrepencies in the measured photocenter in the U, B, V, R, I, J, and H bandpasses, with respect to the K-band photocenter, would be $\sim$0.1, -9.0, -2.0, 1.5, 6.0, 1.0, and 0.2 $\mu$as respectively. Just as in the case of the uncertainity in the value of $\beta_{1}$, this discrepancy between equally good fitting high and low inclination solutions can be resolved if one has three or more wavelengths obtained in NA mode.

As mentioned in \S\ref{introsection}, we note that the limb-darkening function, which was automatically chosen by the ELC code as incorporated into the model atmospheres, can differ from actual observed values by $\sim$10\% \citep{Claret08}. We have tested how changing the limb-darkening coefficients by 10\% affects the resulting astrometric displacements, and find that the result is less than 0.5\% for all wavelengths, and thus is negligible in the modeling. The reason is that limb-darkening is symmetric, and thus while increased limb-darkening damps the visible pole, it also damps the rest of the star, and thus the relative brightness between regions is maintained. 

Additionally, this inclination technique yields the orientation of the projected stellar rotation axis on the sky, which is parallel to the wavelength dispersion direction. When coupled with the derived inclination, this technique thus yields the full 3-dimensional orientation of the rotation axis. This could be a powerful tool in determining the overall alignment of stellar axes in the local neighborhood and in nearby clusters.

\section{Star Spots}
\label{spotsection}

Another area of astrophysical interest to which multi-wavelength astrometric measurements from SIM Lite can contribute is the study of star spots. As the cause of star spots are intense magnetic fields at the photosphere, they are typically found in stars with convective envelopes, especially rapidly rotating stars. Thus, both low-mass, main-sequence K and M dwarfs, as well as rapidly rotating giant and sub-giant stars, are known to host large spots on their surface. The study of the distribution, relative temperature, and size of these spots would greatly contribute to the study of magnetic field generation in stellar envelopes. A starspot that rotates in and out of view will cause a shift of the photocenter for a single star, which has been a subject of much recent discussion in the literature \citep[e.g.][]{Hatzes02,Unwin05,Eriksson07,Catanzarite08,Makarov09,Lanza08}, especially in light of its potential to mimic, or introduce noise when characterizing, an extrasolar planet. However, there has been no mention in the literature of the multi-wavelength astrometric signature of stellar spots, where, just as in the case of the gravity darkening inclination effect, we are looking at essentially two blackbodies with varying temperatures, and thus shorter wavelengths will be more affected by a spot than longer wavelengths.

\begin{figure*}
\centering
\begin{tabular}{cc}
\epsfig{file=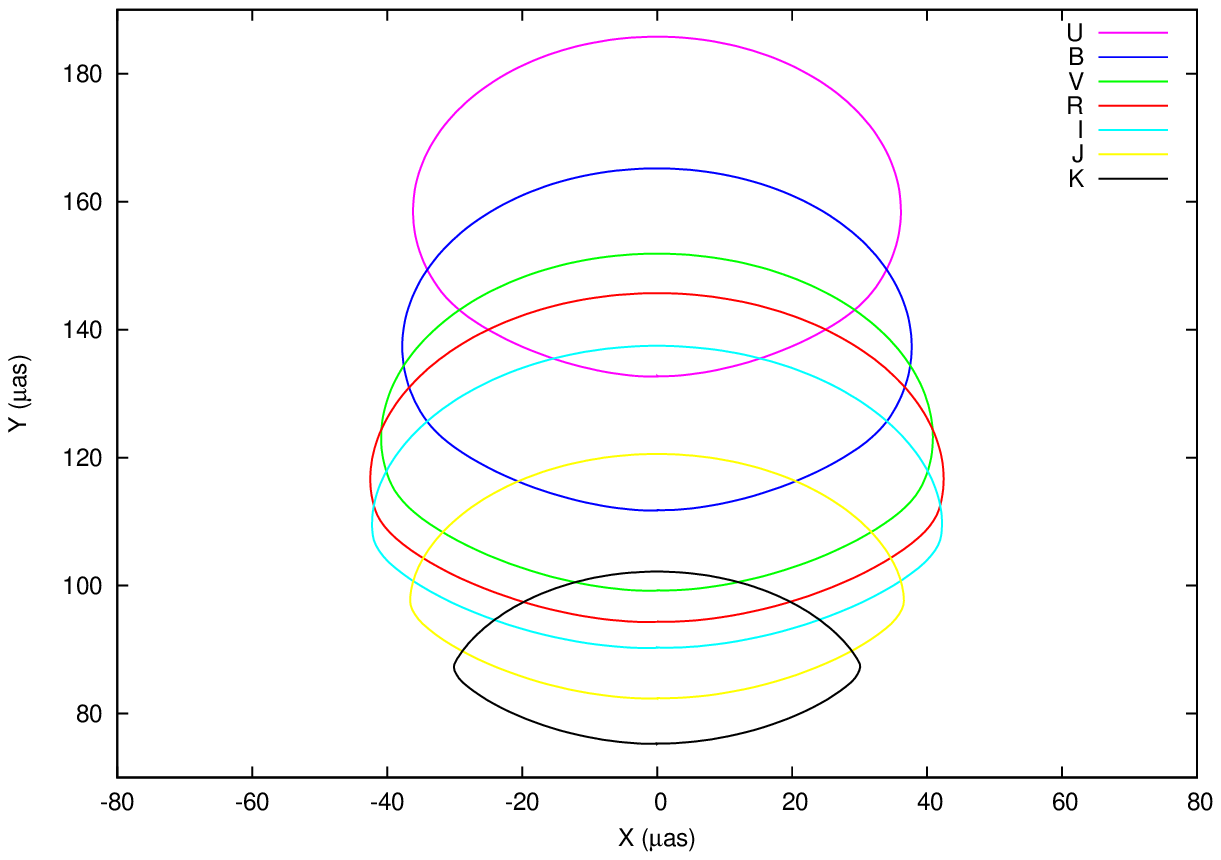, width=0.45\linewidth} &
\epsfig{file=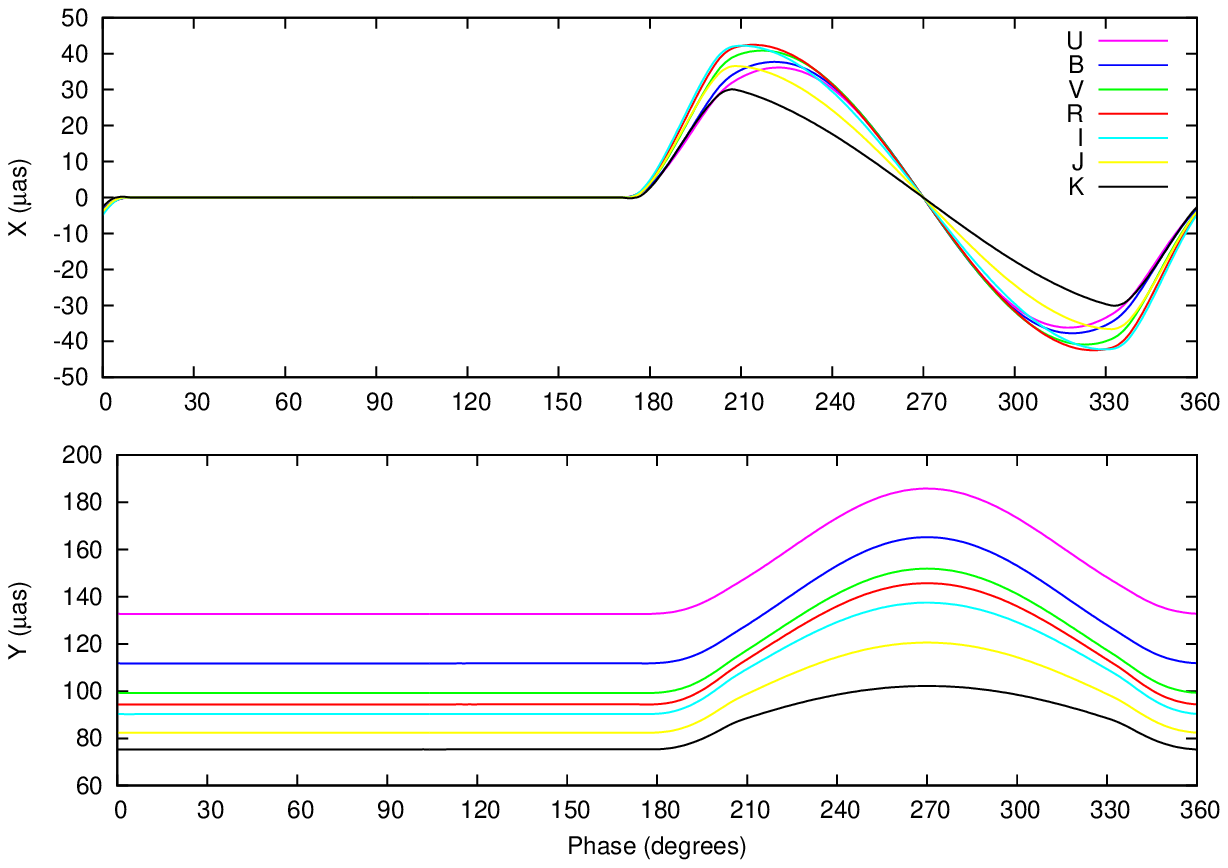, width=0.45\linewidth} \\
\end{tabular}
\caption{A simulated cool Spot on CapellaAb. The spot is located on the equator, with a longitude such that it is seen directly at phase 270$\degr$. The strong presence of the gravity darkening effect, discussed in \S\ref{incsection}, dominates the wavelength spread in the y direction. Note that the broad wavelength covereage of SIM Lite will only cover the B, V, R, and I bandpasses.}
\label{CapellaAbSpotFig}
\end{figure*}

\begin{figure*}
\centering
\begin{tabular}{cc}
\epsfig{file=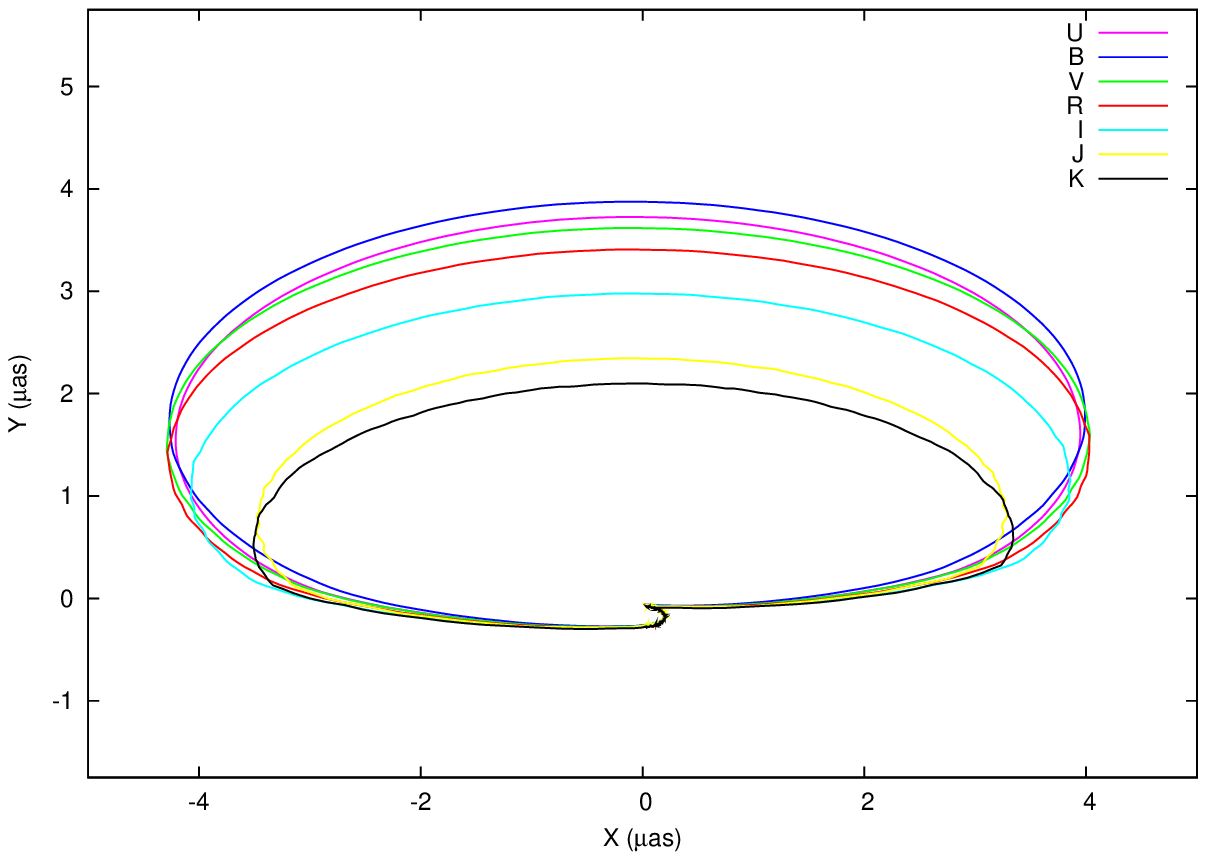, width=0.45\linewidth} &
\epsfig{file=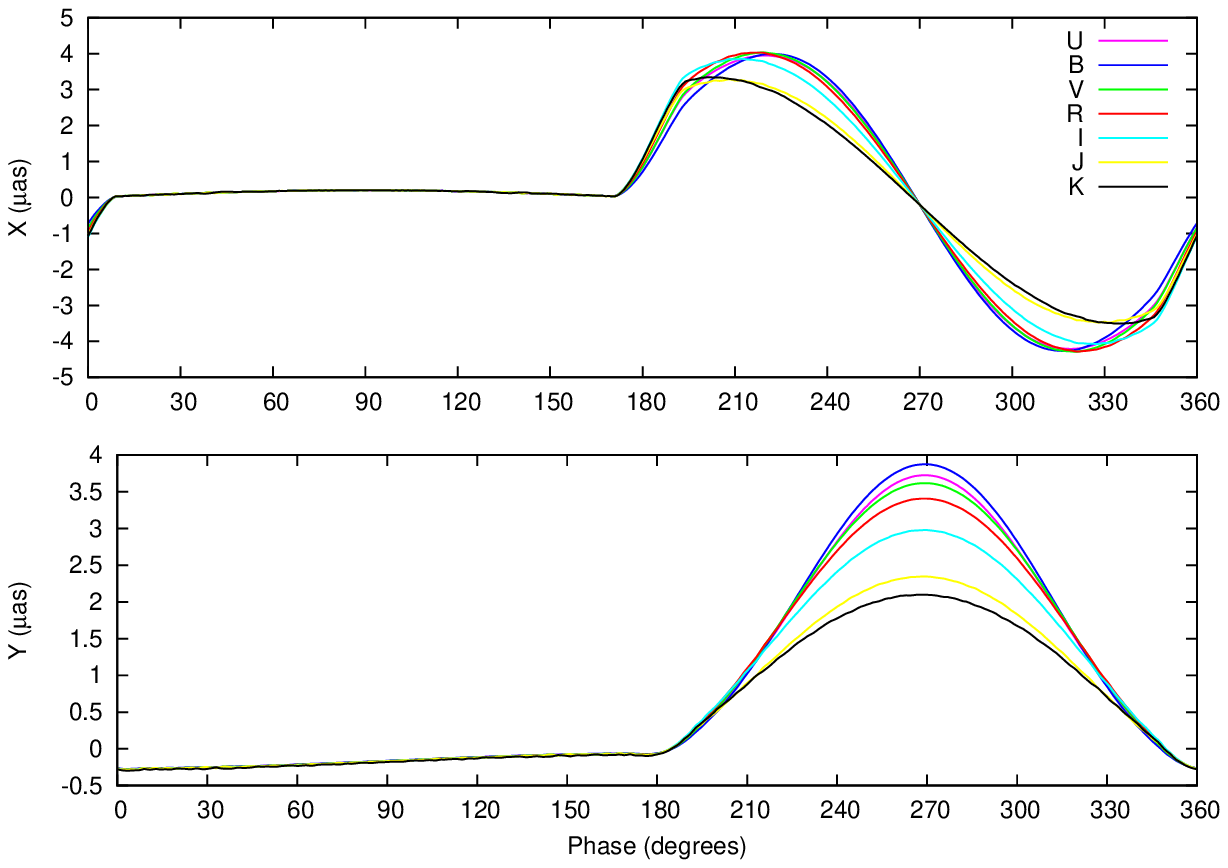, width=0.45\linewidth} \\
\end{tabular}
\caption{A simulated cool spot on a nearby K dwarf star with an inclination of 60$\degr$, whose parameters are given in Table~\ref{kdwarftable}. The spot is located on the equator, with a longitude such that it is seen directly at phase 270$\degr$. Note that the broad wavelength covereage of SIM Lite will only cover the B, V, R, and I bandpasses.}
\label{SingleStarCoolSpot}
\end{figure*}

\begin{figure*}
\centering
\begin{tabular}{cc}
\epsfig{file=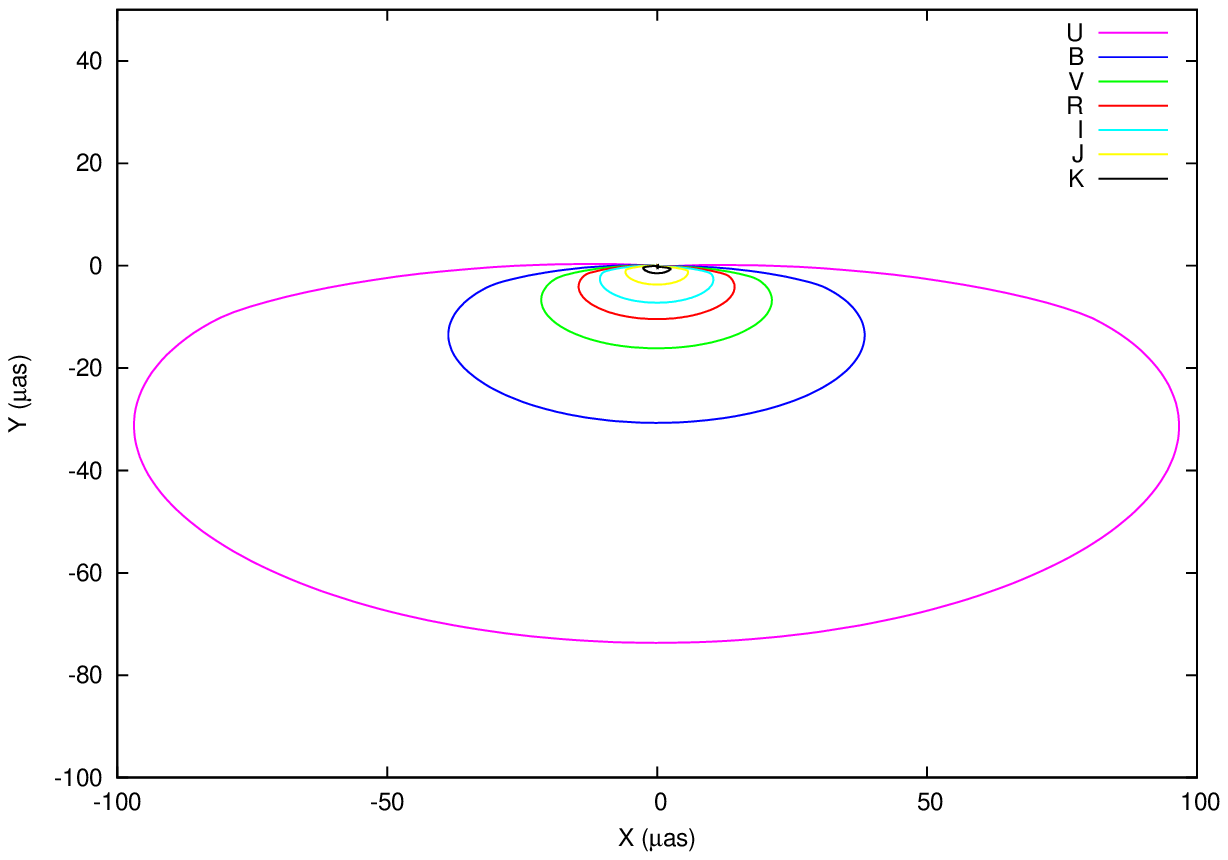, width=0.45\linewidth} &
\epsfig{file=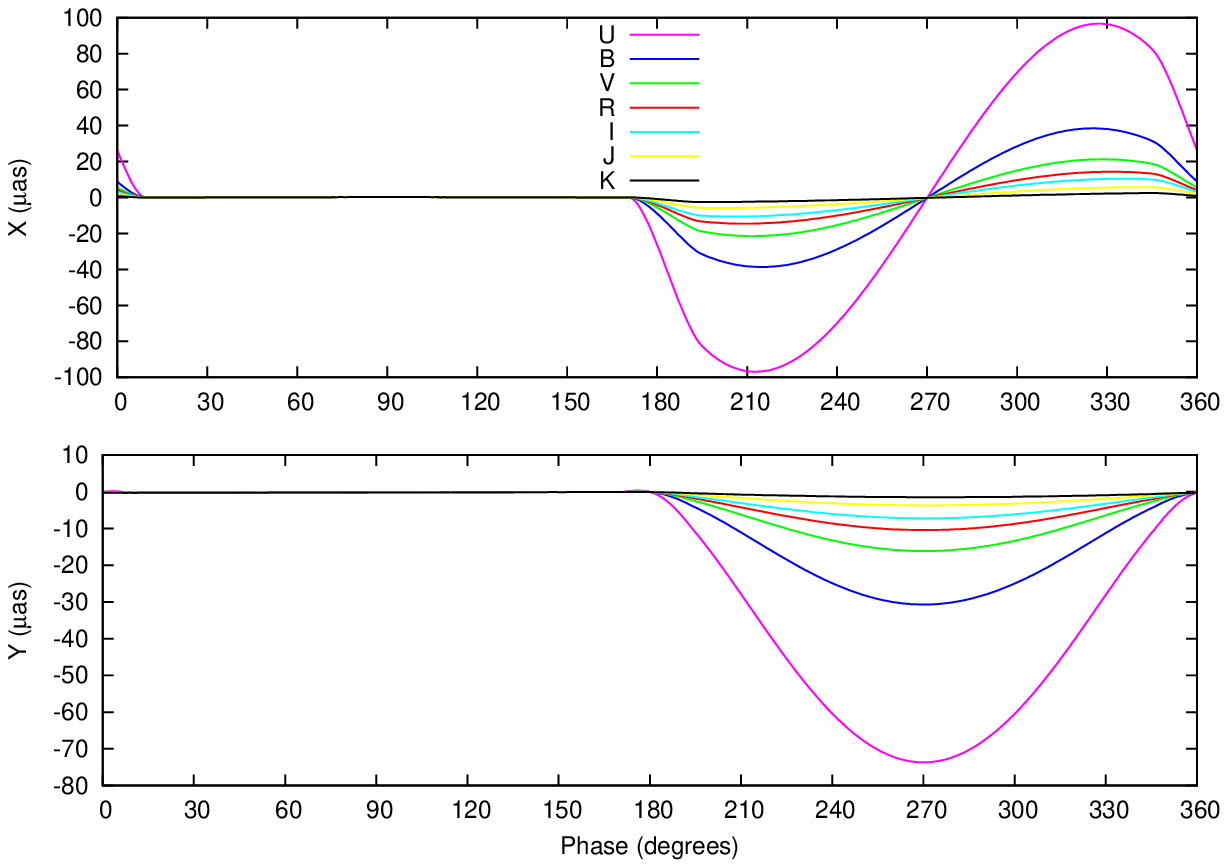, width=0.45\linewidth} \\
\end{tabular}
\caption{A simulated hot spot or flare on a nearby K dwarf star with an inclination of 60$\degr$, whose parameters are given in Table~\ref{kdwarftable}. The spot is located on the equator, with a longitude such that it is seen directly at phase 270$\degr$. Note that the broad wavelength covereage of SIM Lite will only cover the B, V, R, and I bandpasses.}
\label{SingleStarHotSpot}
\end{figure*}

To characterize the multi-wavelength astrometric signature of stellar spots, we model two spotty systems, again using the {\sc reflux} code. We model Capella Ab, which shows evidence of large spots and is suspected of being a RS CVn variable \citep{Hummel94}, and a typical main-sequence K dwarf. For Capella Ab, we use the parameters listed in Table~\ref{CapellaAbTable}, along with the star's determined inclination of 42.788$\degr$ \citep{Torres09}, and add a cool spot that has a temperature that is 60\% of the average surface temperature, located at the equator, at a longitude such that it is seen directly at phase 270$\degr$, and having an angular size of 10$\degr$, (where 90$\degr$ would cover exactly one half of the star). For the K dwarf system, we use the physical parameters listed in Table~\ref{kdwarftable}, simulating a typical K Dwarf at 10 parsecs, and add a cool spot with the same parameters as we do for Capella Ab. Additionally, to investigate the effects of cool versus hot spots or flares, we also run a model with a hot spot by changing the spot temperature to be 40\% greater than the average surface temperature. We present our models in Figures~\ref{CapellaAbSpotFig},~\ref{SingleStarCoolSpot}, and~\ref{SingleStarHotSpot}.

\begin{deluxetable}{lc}
\tablewidth{0pt}
\tablecaption{Parameters for the K Dwarf System}
\tablecolumns{2}
\tablehead{Parameter & Value}
\startdata
Distance (pc) & 10.0\\
Inclination ($\degr$) & 60.0\\
Period (Days) & 20\\
Mass (M$_{\sun}$) & 0.6\\
Radius (R$_{\sun}$) & 0.6\\
Effective Temperature (K) & 4500\\
Latitude of Spot ($\degr$) & 90\\
Longitude of Spot ($\degr$) & 270\\
Angular Size of Spot ($\degr$) & 10\\
Cool Spot Temperature Factor & 0.6\\
Hot Spot Temperature Factor & 1.4
\enddata
\label{kdwarftable}
\end{deluxetable}

As can be seen for CapellaAb, the gravity darkening inclination effect presented in \S\ref{incsection} dominates the spread of colors in the y-direction, the direction parallel to the stars' projected rotation axis. However, the amplitude of the spot motion is quite large, with a total amplitude of $\sim$40 $\mu$as in all bandpasses, which would be easily detectable by SIM Lite. For the K dwarf with a cool spot, we see a much smaller, but still detectable shift of amplitude $\sim$5-8 $\mu$as, depending on the wavelength. In the case of a hot spot or flare, we see a much larger displacement, on the order of $\sim$10-200 $\mu$as, depending on the wavelength, which would be easily detectable by SIM and provide extremely precise values in deriving the spot parameters. 

In general, the temperature of the spot, in relation to the mean stellar surface temperature, is related to the spread in observed wavelengths, with a larger spread indicating a larger temperature difference. The duration of the astrometric displacement in phase, coupled with the overall amplitude of the astrometric displacement, yields the size of the spot, as larger spots will cause larger displacements and be visible for a larger amount of rotational phase. The latitude of the spot can also affect the total duration. Finally, the amplitude of the astrometric displacement in the x versus the y direction is dependent on both the latitude of spot as well as the inclination of the star. Thus, when modeled together, one is able to recover these parameters. This work can also be combined with our work in Paper I to derive the location of spots in binary systems, as the astrometric signature of the spot is simply added to the astrometric signature of the binary system.

The astrometric motion induced upon a parent star by a host planet does not have a wavelength dependence. Spots however, as we have shown via our modeling, have a clear wavelength dependence. Thus, if one has a candidate planetary signal from astrometry, but it shows a wavelength-dependant motion, it must then be a false positive introduced from star spots at the rotation period of the star, (assuming that the planet's emitted flux is negligible compared to the star.) Furthermore, when SIM is launched, there will likely be many cases where a marginally detectable signal due to a planetary companion is found at a very different period than the rotation period of the star. However, starspots will still introduce extra astrometric jitter which will degrade the signal from the planetary companion. Multi-wavelength astrometric data can be used to model and remove the spots, which will have a wavelength dependence, and thus strengthen the planetary signal, which will not have a wavelength dependence.

\section{Discussion and Conclusion}
\label{conclusionsection}

We have presented detailed models of the multi-wavelength astrometric displacement that SIM Lite will observe due to gravity darkening and stellar spots using the \textsc{reflux} code. We find that SIM Lite observations, especially when combined with other techniques, will be able to determine the absolute inclination, gravity darkening exponent, and 3-dimensional orientation of the rotational axis for fast and slow rotating giant stars, and fast-rotating main-sequence stars. This technique will be especially useful in probing binary star and exoplanet formation and evolution, as well as the physics of star forming regions. Direct observational determination of the gravity darkening exponent has direct applications in both stellar and exoplanet astrophysics. This technique is also relatively inexpensive in terms of SIM Lite observing time, as one need only to observe a given star once, as opposed to binary stars and planets, which require constant monitoring over an entire orbit. It should be noted that this effect should be taken into account when constructing the SIM Lite astrometric reference frame, such that fast-rotating giants should be excluded so as not to produce a wavelength-dependent astrometric reference fame.

We also have presented models of star spots on single stars, and find that SIM Lite should be able to discern their location, temperature, and size. Combined with other techniques, this will provide great insight into stellar differential rotation, magnetic cycles and underlying dynamos, and magnetic interaction in close binaries. From this modeling, it should especially be noted that multi-wavelength astrometry is a key tool in the hunt for extrasolar planets, either by ruling out false signals created by spots, or simply removing extra astrometric jitter introduced by spots. Thus, it remains critical that SIM Lite maintains a multi-wavelength astrometric capability in its final design.

\acknowledgments
This work was sponsored in part by a SIM Science Study (PI: D. Gelino) from the National Aeronautics and Space Administration through a contract with the Jet Propulsion Laboratory, California Institute of Technology. J.L.C. acknowledges additional support from a New Mexico Space Grant Consortium Fellowship. We thank the referee for comments which greatly helped to improve this manuscript, particularly in making the presentation of our ideas much clearer.

\bibliography{refs.bib}

\end{document}